\documentclass[english]{IEEEtran}
\usepackage[T1]{fontenc}
\usepackage[latin9]{inputenc}
\usepackage{amsmath}
\usepackage{graphicx}

\makeatletter

\newcommand{\lyxmathsym}[1]{\ifmmode\begingroup\def\b@ld{bold}
  \text{\ifx\math@version\b@ld\bfseries\fi#1}\endgroup\else#1\fi}

\providecommand{\tabularnewline}{\\}

\usepackage{hyperref}

\@ifundefined{showcaptionsetup}{}{%
 \PassOptionsToPackage{caption=false}{subfig}}
\usepackage{subfig}
\makeatother

\usepackage{babel}
\begin{document}
\title{Proof of Useful Intelligence (PoUI): Blockchain Consensus Beyond Energy
Waste}
\author{%
\begin{minipage}[t]{0.3\textwidth}%
\begin{center}
Zan-Kai Chong\\
\emph{School of Science and\\ Technology\\
Kwansei Gakuin University}\\
Japan\\
zankai@ieee.org
\par\end{center}%
\end{minipage}%
\begin{minipage}[t]{0.3\textwidth}%
\begin{center}
Hiroyuki Ohsaki\\
\emph{School of Science and\\ Technology\\
Kwansei Gakuin University}\\
Japan\\
ohsaki@kwansei.ac.jp
\par\end{center}%
\end{minipage}%
\begin{minipage}[t]{0.3\textwidth}%
\begin{center}
Bryan Ng\\
\emph{ 
School of Engineering \& Computer Science\\
Victoria University of Wellington}\\
New Zealand
\\
bryan.ng@ecs.vuw.ac.nz
\par\end{center}%
\end{minipage}}
\maketitle
\begin{abstract}
Blockchain technology anchors decentralized systems by enabling secure,
transparent data management across distributed networks, powering
a wide range of applications --- from foundational cryptocurrencies
like Bitcoin to the recently emerging tokenization of real-world assets
(RWAs), such as property and commodities. However, its scalability
and environmental sustainability depend on consensus mechanisms that
maintain network integrity without imposing excessive computational
or energy burdens. Proof of Work (PoW), a prevalent mechanism seen
in Bitcoin, relies on miners performing energy-intensive cryptographic
computations to ensure robust security, yet driving significant resource
demands. In contrast, Proof of Stake (PoS) selects validators based
on the amount of cryptocurrency they stake, as exemplified by Ethereum
post-Merge, providing a markedly more energy-efficient option than
PoW. While PoW excels in delivering decentralized security through
computational effort, it does so at the cost of high energy consumption;
PoS, meanwhile, enhances participation accessibility and reduces resource
use but introduces potential centralization risks due to wealth concentration
among larger stakers. The rapid rise of artificial intelligence (AI)
models, with their substantial energy consumption, underscores a growing
strain on computational resources. Hence, it inspires us to propose
a new consensus mechanism, namely, Proof of Useful Intelligence (PoUI).
PoUI is a hybrid consensus mechanism where workers execute AI-based
tasks, such as natural language processing or image analysis, to earn
coins, which are then staked to secure the network, seamlessly integrating
security with real-world utility. This system leverages decentralized
functional nodes, i.e., job posters who submit tasks, market coordinators
who oversee jobs distribution, workers who perform computations, and
validators who ensure accuracy, all orchestrated by smart contracts
for task execution and reward allocation. Our energy analysis benchmarks
PoW at 3.51 kWh/miner, PoS at 0.1 kWh/validator, and PoUI at 0.6 kWh/worker
--- yielding a 97\% energy reduction from PoW while adding value.
Simulations further demonstrate that PoUI\textquoteright s dynamic
reward adjustment regulates worker participation in the job market,
which subsequently encourages a sufficient number of validators in
the network.

\end{abstract}

\section{Introduction}

Blockchain technology has revolutionized decentralized systems by
enabling secure, intermediary-free transactions, fundamentally reshaping
trust in digital environments \cite{nakamoto2008,reyna2018blockchain}.
Introduced by Satoshi Nakamoto in 2008 as Bitcoin\textquoteright s
foundation, blockchain is a distributed ledger where transactions
are cryptographically linked into blocks, replicated across a network
of nodes, and secured without centralized control. This architecture
ensures immutability and transparency, driving applications beyond
cryptocurrencies such as supply chain traceability and digital identity
management, to enhance accountability and efficiency. One of the blockchain's
recent applications is the tokenization of real-world assets (RWAs),
like real estate and art, which leverages blockchain to fractionalize
ownership, boosting liquidity and democratizing access to high-value
markets \cite{cong2021tokenomics}.

The stability of blockchain networks hinges on consensus mechanisms,
i.e., protocols that ensure network agreement, but their energy demands
pose significant challenges. Proof of Work (PoW) is the Bitcoin\textquoteright s
core mechanism that exemplifies this issue. It relies on energy-intensive
cryptographic computations, which are estimated to be at 181.67 terawatt-hours
(TWh) annually in 2025 \cite{CBECI2024} to secure the network ---
a burden that continues to spark environmental concerns \cite{o2014bitcoin,deVries2022,Stoll2019CarbonFootprintBitcoin}.
PoW diverts vast resources to puzzles offering no practical utility
beyond security. As blockchain adoption grows across diverse domains,
the inefficiencies of PoW have intensified the need for sustainable
consensus alternatives that balance security with tangible benefits
\cite{buterin2020incentives}.

The rapid emergence of artificial intelligence (AI) models, such as
Large Language Models (LLMs) for text and image generation, has introduced
a parallel challenge: their substantial power consumption \cite{samsi2023words}.
Running LLMs demand vast computational resources, often exceeding
hundreds of megawatt-hours per model, rivaling the energy intensity
of PoW-based blockchains. This confluence of energy-hungry technologies
inspires us to reimagine PoW\textquoteright s wasteful computations.
Hence, we propose Proof of Useful Intelligence (PoUI), a consensus
mechanism that leverages AI models to perform valuable AI tasks while
securing the blockchain, transforming PoW\textquoteright s inefficiencies
into productive outcomes.

At the core of PoUI is a decentralized marketplace that orchestrates
task distribution and execution across the network. This marketplace
involves four key roles: market coordinators, nodes with sufficient
stake and reputation that manage job clusters and match tasks with
workers; job posters, who submit tasks to the marketplace; workers,
who execute tasks using AI models; and validators, who ensure the
integrity of communications and task outcomes. Tasks are designed
to be divisible into smaller, manageable chunks to enable flexible
participation without real-time constraints. Upon completion, smart
contracts automate payments, transferring compensation from job posters
to both market coordinators and workers, incentivizing engagement
and ensuring fairness.

PoUI re-imagines blockchain consensus as a dual-purpose system that
secures the network and creates tangible value through AI. This paper
is structured as follows. In Section \ref{sec:Related-Work}, we explore
traditional PoW and PoS mechanisms alongside recent advancements in
Proof of Useful Work (PoUW) and Proof of Intelligence (PoI). Next,
in Section \ref{sec:PoUI-Network-Architecture}, we delve into the
roles of functional nodes, illustrated with a simplified example.
We then compare the energy consumption of PoW, PoS, and PoUI in Section
\ref{sec:Energy-Consumption-Analysis} with some assumptions. Following
this, Section \ref{sec:PoUI-Market-Mechanism} discusses the market
mechanism designed to regulate the number of workers effectively in
processing the jobs with some simulation results. Finally, we present
our conclusions in Section \ref{sec:Conclusion}, underscoring PoUI\textquoteright s
role as a pioneering step toward socially impactful blockchain technology.

\section{Related Work \label{sec:Related-Work}}

Consensus mechanisms are critical to blockchain networks, enabling
decentralized nodes to agree on the state of the ledger while ensuring
security and trust. Various mechanisms have been developed, each balancing
trade-offs in energy use, decentralization, and utility. This section
examines the prominent consensus mechanisms with our proposed Proof
of Useful Intelligence (PoUI).

\subsection{Proof of Work (PoW)}

PoW is the consensus mechanism used in Bitcoin, and it requires miners
to solve computationally intensive cryptographic puzzles to validate
transactions and add blocks to the blockchain. The first miner to
solve the puzzle is rewarded, securing the network through this competitive
process. The key advantages of PoW are its proven security as it relies
on computational effort to make it highly resistant to attacks. Additionally,
the mechanism is straightforward and widely understood.

However, PoW has significant drawbacks. The vast computational power
required consumes enormous amounts of energy, often with no value
beyond network security, raising environmental concerns. Additionally,
mining has increasingly concentrated in the hands of large entities
with specialized hardware, undermining decentralization. In contrast,
PoUI offers significant energy efficiency, using only 0.6 kWh per
node versus PoW\textquoteright s 3.51 kWh, while generating valuable
AI-driven outputs (see Section \ref{sec:Energy-Consumption-Analysis}).
It also promotes broader participation through its job market, reducing
centralization compared to PoW\textquoteright s hardware-driven model.

\subsection{Proof of Stake (PoS)}

PoS offers an alternative by selecting validators based on the amount
of cryptocurrency they hold and stake, rather than their computational
power. This reduces the need for energy-intensive calculations as
compared to PoW. Also, it potentially broadens accessibility by eliminating
the need for expensive hardware.

Yet, PoS has its own limitations. Validators with larger stakes are
more likely to be chosen, which can centralize power among the wealthiest
participants. Issues like the \textquotedbl nothing-at-stake\textquotedbl{}
problem, where validators support multiple chains, pose new vulnerabilities.
By contrast, PoUI extends beyond PoS\textquoteright s security-only
framework, employing AI models to deliver practical outcomes like
text and image generation. Its marketplace structure fosters inclusivity
by compensating a wide range of participants, countering PoS\textquoteright s
tendency toward wealth concentration (see Section \ref{sec:PoUI-Network-Architecture}).

\subsection{Proof of Useful Work (PoUW) and Variations}

PoUW is first introduced in \cite{lihu2001proof} with the aim to
replace the computationally wasteful PoW used in systems like Bitcoin.
The paper outlines a system where the computational efforts of miners
are redirected toward training machine learning (ML) models, thereby
producing socially beneficial outcomes rather than merely solving
cryptographic puzzles.

Generally, the miners compete to train ML models, submitting their
solutions to a smart contract that evaluates the quality of the work
based on predefined metrics, such as model accuracy or loss on a validation
dataset. Then, the miner producing the best model wins the block reward,
consisting of both the client\textquoteright s payment and standard
blockchain incentives.

Some variations of PoUW are observed in different implementations.
For instance, \cite{dong2019proofware} leverages computational power
for decentralized applications (dApps), while \cite{baldominos2019coin}
focuses on training deep learning models, requiring model performance
to exceed a predefined threshold for block validation. Additionally,
\cite{baldominos2019coin} applies PoUW to combinatorial optimization
problems, such as scheduling and vehicle routing. However, these implementations
have notable limitations. The approaches in \cite{lihu2001proof}
and \cite{baldominos2019coin} are constrained to ML and deep learning
tasks, demanding specialized hardware and expertise, which restricts
participation to well-resourced nodes. Similarly, \cite{baldominos2019coin}
targets niche optimization problems, limiting its applicability and
contributor pool. While \cite{dong2019proofware} supports a broader
range of dApps, it lacks a cohesive framework for task allocation
and reward distribution, reducing its scalability and efficiency.

In contrast, our proposed PoUI addresses these shortcomings by introducing
a decentralized job market that supports a diverse range of AI-driven
tasks, from natural language processing to public goods contributions
(e.g., verifying open datasets). Unlike the specialized focus of \cite{lihu2001proof},
\cite{baldominos2019coin}, and \cite{todorovic2022proof}, PoUI leverages
widely accessible AI models, lowering barriers to entry and broadening
participation. Furthermore, PoUI\textquoteright s market coordinators
and smart contract-based task management provide the unified coordination
absent in \cite{dong2019proofware}, ensuring efficient task distribution
and reward allocation. By integrating a PoS-based validator selection
with dynamic reward adjustments, PoUI enhances inclusivity and scalability,
offering a more versatile and socially impactful consensus mechanism
compared to existing PoUW implementations.

\subsection{Proof of Intelligence (PoI)}

PoI is a consensus mechanism that leverages computational resources
for AI-related tasks, such as neural network inference, to secure
the blockchain while producing valuable outputs \cite{cortex2018}.
In PoI, nodes compete to execute predefined AI tasks, with the best-performing
node (e.g., based on model accuracy) earning the right to validate
the block. This approach reduces energy waste compared to PoW by redirecting
computational effort toward practical applications.

PoI\textquoteright s task scope, however, is often limited to specific
ML frameworks. It requires specialized hardware and expertise, which
may restrict participation and scalability. Additionally, its consensus
relies solely on task performance, potentially favoring computationally
powerful nodes and risking centralization. In contrast, our proposed
PoUI introduces a decentralized job market supporting diverse AI-driven
tasks, from text generation to public goods contributions, and integrates
a PoS-based validator selection to enhance inclusivity and decentralization
(see Section \ref{sec:PoUI-Network-Architecture}).

\section{PoUI Network Architecture \label{sec:PoUI-Network-Architecture}}

In the following, we elaborate on the architecture of the PoUI network
and the corresponding functional nodes.

\subsection{Functional Nodes in the PoUI Network}

The PoUI consensus mechanism operates through four types of functional
nodes, each performing distinct roles to support the network\textquoteright s
operation, security, and utility. These functional nodes are:
\begin{itemize}
\item Job Posters: Nodes that submit tasks to the network for execution.
\item Market Coordinators: Nodes responsible for coordinating job postings
and managing the work completed by workers.
\item Workers: Nodes that perform the tasks using AI models such as LLMs
or similar computational resources.
\item Validators: Nodes responsible for validating the work and adding blocks
to the blockchain.
\end{itemize}
In the following, we detail the role and operation of each functional
node within the PoUI ecosystem.

\subsubsection{Job Posters}

Job posters are functional nodes that initiate the process by submitting
tasks to the job market, which is managed by market coordinators.
These tasks require computational services, typically provided by
AI models, such as text and image generation and may even contribute
to public goods (e.g., verifying Wikipedia entries).

Tasks are categorized into two types, i.e., private jobs, which benefit
the job poster directly, and public good jobs, which provide value
to the broader community. For private jobs, job posters are typically
required to pay a fee in the network\textquoteright s currency to
incentivize workers and cover operational costs. Conversely, for public
good jobs, this payment may be waived, with rewards potentially subsidized
by the network (e.g., through block rewards or a community fund) to
encourage contributions to societal benefit. Each job submission includes
detailed specifications, such as:
\begin{itemize}
\item Job Type: The category of task (e.g., image processing).
\item Job Description: A detailed explanation of the task requirements. 
\item Job Validity Period: The duration for which the job remains available
before it is removed from the market if not accepted. 
\item Runtime Requirements: The estimated computational time or resources
needed tocomplete the task once accepted.
\end{itemize}

\subsubsection{Market Coordinators}

Market coordinators are the functional nodes that manage the job market
and facilitate interactions between job posters and workers. When
a job poster submits a task, its validity is evaluated by the coordinators;
if accepted, the task is added to the job queue --- a decentralized
list of available tasks. Job queues are synchronized across all coordinators
to ensure consistency in the network\textquoteright s task list. The
history of job posters, including metrics like submission frequency
and task quality, is recorded and tracked by coordinators to maintain
reliability and prevent abuse.

\subsubsection{Workers}

Workers are functional nodes equipped with AI models or similar computational
resources, tasked with executing the jobs listed by market coordinators.
Workers review the job listings and select tasks they are interested
in and capable of completing.

When a worker accepts a job, a smart contract is established among
the market coordinator, the job poster, and the worker. This smart
contract governs the task execution process, including deadlines,
quality requirements, and reward distribution. Upon successful completion
of the tasks, both workers and market coordinators earn coins as rewards.

While it is possible for a group of workers to form a cluster to secure
and execute jobs collaboratively, similar idea on collaborative computation
has been found in \cite{chong2025llmnetdemocratizingllmsasaserviceblockchainbased,Shafay2022BlockchainDL}
and will not be discussed in this paper.

\subsubsection{Validators}

Validators are functional nodes responsible for ensuring the integrity
of the work performed by workers and securing the blockchain by adding
validated blocks. Validators are selected using PoS mechanism, where
nodes with higher stakes have a greater likelihood of being chosen.
Validators perform a critical role in verifying the quality and correctness
of workers\textquoteright{} outputs before these are included in the
blockchain. Notably, a validator can also act as a worker, enabling
flexible participation when the demand for workers is high. However,
it cannot validate its own work to avoid conflicts of interest and
preserve trust.

\subsection{Centralization Risks and Mitigation \label{subsec:Centralization-Risks}}

The PoS-based validator selection risks centralization if active workers
accumulate disproportionate coins to increase their stake and influence.
To address this, we adopt existing methods such as (i) applying stake
caps to limit the maximum coins stakable per validator, preventing
dominance by high-earners \cite{kiayias2017ouroboros}; and (ii) adding
random selection adjustments to give less-staked nodes better odds
\cite{buterin2020incentives}. These measures ensure balanced participation
and preserve decentralization across the validator pool.

\subsection{Security Against Malicious Attacks}

AI-driven tasks in PoUI, such as text or image generation, lack a
single correct answer, making the network vulnerable to malicious
attacks. These include fraudulent tasks by job posters, low-quality
outputs by workers, or collusion between nodes to approve invalid
work. PoUI mitigates these risks with the following measures:
\begin{enumerate}
\item Task Screening: Market coordinators check submitted tasks for legitimacy
based on job poster history and task relevance, rejecting suspicious
tasks to prevent fraud.
\item Output Verification: Validators use frameworks like MCP \cite{anthropic2024mcp}
or CodeAct \cite{wang2024executable} to assess AI output quality.
Multiple validators review subjective tasks, requiring majority agreement
for approval to counter low-quality submissions.
\item Collusion Prevention: Validators are randomly assigned to tasks and
cannot verify their own work. Smart contracts log all decisions transparently
to detect collusion.
\item Reputation System: Nodes earn reputation scores based on honest behavior.
Malicious actions, like submitting fraudulent tasks or invalid outputs,
lower scores, reducing rewards or access to tasks.
\end{enumerate}
These strategies ensure PoUI\textquoteright s resilience against malicious
attacks, maintaining network trust. Future work will explore advanced
attack detection to further enhance security.

\subsection{Incentive Structure and Dynamic Reward Balancing}

In PoUI, workers complete jobs from job market to earn coins. These
coins can then be staked by workers to qualify as validators under
the PoS mechanism, creating a pathway for active contributors to gain
influence in the network. Besides, the workers and validators may
sell the coins to the job posters to fund new tasks or exchange for
other resources, enhancing liquidity within the ecosystem.

To ensure a balanced ecosystem, the rewards earned by workers are
dynamically adjusted based on job demand (see Section \ref{sec:PoUI-Market-Mechanism}).
Specifically, when the number of pending jobs in the market exceeds
a predefined threshold, workers\textquoteright{} rewards are increased
to incentivize task completion and clear the backlog. Conversely,
when the job queue is low, rewards for workers are reduced, reflecting
lower demand. This dynamic adjustment mechanism ensures the network
adapts efficiently to varying workloads, encouraging workers to remain
active and maintain system throughput.

\subsection{Example}

\begin{figure}
\begin{centering}
\subfloat[]{\begin{centering}
\includegraphics[width=0.4\columnwidth]{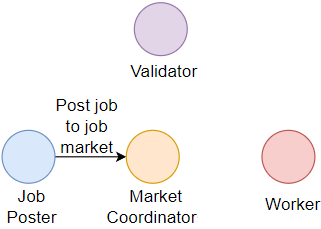}
\par\end{centering}

}\subfloat[]{\begin{centering}
\includegraphics[width=0.4\columnwidth]{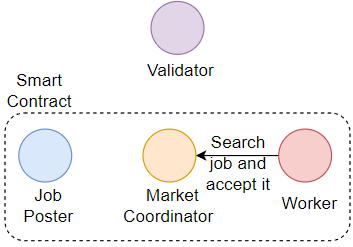}
\par\end{centering}
}
\par\end{centering}
\begin{centering}
\subfloat[]{\begin{centering}
\includegraphics[width=0.4\columnwidth]{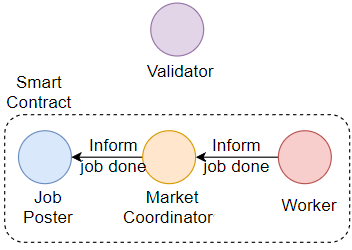}
\par\end{centering}
}\subfloat[]{\begin{centering}
\includegraphics[width=0.4\columnwidth]{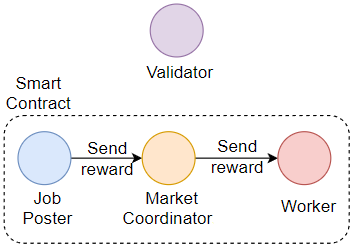}
\par\end{centering}
}
\par\end{centering}
\caption{Workflow of the PoUI consensus mechanism: (a) Job poster submits a
task to market coordinator, (b) Worker searches and accepts a job,
establishing a smart contract, (c) Worker informs market coordinator
of job completion, validated by validator, (d) Job poster sends rewards
via smart contract to market coordinators and worker. \label{fig:workflow}}

\end{figure}

Fig. \ref{fig:workflow} depicts the simplified workflow of the PoUI
consensus mechanism across four key stages, i.e., job posting, job
acceptance, job completion, and reward distribution.

In the first stage, as shown in Fig. \ref{fig:workflow}(a), a job
poster initiates the process by submitting a task to the job market.
The job poster sends the job details, including specifications such
as job type, description, validity period, and runtime requirements,
to the market coordinator. The market coordinator, responsible for
managing the job market, evaluates the task\textquoteright s validity
and, if accepted, adds it to a decentralized job queue, making it
available for a worker to access.

Once the job is listed, the second stage, depicted in Fig. \ref{fig:workflow}(b),
involves a worker searching for and accepting a job. The worker interacts
with the market coordinator to browse the job queue, selects a task
based on its interest and capability, and accepts it. Upon acceptance,
a smart contract is established among the job poster, market coordinator,
and worker. This smart contract governs the task execution process,
enforcing deadlines, quality requirements, and reward distribution.

In the third stage, shown in Fig. \ref{fig:workflow}(c), the worker
completes the assigned task and informs the market coordinator of
the job\textquoteright s completion. The market coordinator then notifies
the job poster that the task is done. Throughout this process, validator
oversees the interactions, ensuring the integrity of the communications
and verifying the quality of the worker\textquoteright s output before
it is recorded on the blockchain.

Finally, in the fourth stage, illustrated in Fig. \ref{fig:workflow}(d),
the reward distribution occurs. Upon validation of the completed task,
the smart contract triggers the payment process. The job poster sends
the reward to the market coordinator, who then distributes the coins
to the worker for task execution and retains a portion as a coordination
fee.

\subsection{Practical Issues in Smart Contract Integration and Development}

Integrating PoUI into existing smart contracts or developing new ones
presents practical challenges due to its unique AI-driven task management
and node interactions. Below, we outline the key challenges to the
implementation.
\begin{itemize}
\item Integration with Existing Smart Contracts: Existing platforms like
Ethereum use Solidity for smart contracts, which may not support PoUI\textquoteright s
job market or reputation system natively. A middleware layer, such
as an oracle or off-chain coordinator, could bridge PoUI\textquoteright s
task validation (e.g., using MCP) with on-chain execution, but this
increases latency and costs. Additionally, PoUI\textquoteright s multi-node
interactions (e.g., task screening, output verification) require complex
contract logic may cause gas fees to increase. Optimizing contract
functions, such as batching validator approvals, potentially mitigate
this issue.
\item New Smart Contracts: New contracts must encode PoUI\textquoteright s
dynamic reward adjustments (Section V) and reputation system (Section
III.C). Subjective AI output validation requires multiple validator
inputs, complicating contract logic. Besides, managing a decentralized
job queue and validating AI tasks on-chain could overload networks
with high transaction volumes. Off-chain computation for task screening
or output scoring, with results hashed on-chain, may improve scalability.
\item Common Challenges: Contracts must enforce anti-collusion and reputation
penalties robustly to prevent exploits. It requires rigorous audit
and formal verification. PoUI contracts need to interact with external
AI models or data (e.g., for public good tasks), necessitating secure
oracles or APIs, which adds complexity.
\end{itemize}
These practical considerations highlight the trade-offs in deploying
PoUI. While integration with existing systems offers faster adoption,
new contracts enable tailored functionality. Future work will explore
optimized contract designs and layer-2 integrations to enhance PoUI\textquoteright s
scalability and cost-efficiency.

\section{Energy Consumption Analysis of PoW, PoS, and PoUI \label{sec:Energy-Consumption-Analysis}}

As blockchain technology scales, the energy efficiency of the consensus
mechanisms grows increasingly important. We evaluate the energy consumption
of each consensus mechanism through the following equation,

\begin{equation}
E_{\text{tot}}=E_{\text{sec}}+E_{\text{use}},\label{eq:total-energy}
\end{equation}
 where $E_{\text{sec}}$ represents the energy essential for upholding
blockchain consensus and integrity, while $E_{\text{use}}$ represents
the energy devoted to computations that produce valuable outputs beyond
the essential task of securing the network.

To provide a standardized reference for the energy consumption analysis,
we present the energy characteristics of the Bitmain Antminer S21
Pro \cite{bitmainshop2025}, a modern mining rig used in PoW-based
blockchains like Bitcoin in Table \ref{tab:energy-bitmain-antminer}.
We now explore the energy consumption of PoW, PoS, and PoUI using
Eq. \ref{eq:total-energy}.

\begin{table}
\caption{Energy characteristics of Bitmain Antminer S21 Pro for PoW.\label{tab:energy-bitmain-antminer}}

\centering{}%
\begin{tabular}{|c|c|c|}
\hline 
Parameter & Value & Unit\tabularnewline
\hline 
\hline 
Hash Rate, $H$ & 234 & TH/s\tabularnewline
\hline 
Energy per Hash, $E_{\text{hash}}$ & 15 & J/TH\tabularnewline
\hline 
Power Consumption & 3510 & W\tabularnewline
\hline 
Energy per Hour, $E_{\text{total}}^{\text{PoW}}$ & 3.51 & kWh/miner\tabularnewline
\hline 
\end{tabular}
\end{table}

\subsection{Proof of Work (PoW) \label{subsec:PoW-energy}}

PoW secures the network by leveraging miners to solve complex cryptographic
puzzles, a process that demands substantial energy expenditure. The
security energy per miner over one hour is modeled as,

\begin{equation}
E_{\text{sec}}^{\text{PoW}}=H\times E_{\text{hash}}\times3600,\label{eq:PoW-energy}
\end{equation}
where, $H$ is the hash rate per miner (hashes per second), $E_{\text{hash}}$
is the energy per hash (joules per hash), and 3600 is the number of
seconds in an hour. Since PoW performs no useful work beyond security,
$E_{\text{use}}^{\text{PoW}}=0$, and thus the energy per miner per
hour is $E_{\text{total}}^{\text{PoW}}=E_{\text{sec}}^{\text{PoW}}$.

Using the Bitmain Antminer S21 Pro specification from Table \ref{tab:energy-bitmain-antminer},
the security energy is,

\begin{align}
E_{\text{sec}}^{\text{PoW}} & =(234\times10^{12})\times(15\times10^{-12})\times3600\nonumber \\
 & =3.51\,\text{kWh/miner.}
\end{align}
Thus, $E_{\text{total}}^{\text{PoW}}=3.51\,\text{kWh/miner},$ reflecting
PoW\textquoteright s high energy demand compared to alternative mechanisms.

\subsection{Proof of Stake (PoS)\label{subsec:PoS-energy}}

PoS secures the network by selecting validators based on staked cryptocurrency.
The security energy per validator over one hour is modeled as,

\begin{equation}
E_{\text{sec}}^{\text{PoS}}=P_{\text{val}}\times t,\label{eq:PoS-energy}
\end{equation}
where $P_{\text{val}}$ is the power consumption per validator (watts),
$t$ is time in hours. Since PoS performs no useful work beyond security,
$E_{\text{use}}^{\text{PoS}}=0$, and thus the total energy per validator
over one hour is,

\begin{equation}
E_{\text{tot}}^{\text{PoS}}=E_{\text{sec}}^{\text{PoS}}.
\end{equation}
Assuming $P_{\text{val}}=100\,\text{W}=0.1\,\text{kW}$ for an Ethereum-like
validator node (e.g., a standard PC) and and $t=1\,\text{hour}$,

\begin{align}
E_{\text{sec}}^{\text{PoS}} & =0.1\,\text{kW}\times1\,\text{hour}\nonumber \\
 & =0.1\,\text{kWh/validator}.\label{eq:PoS-energy-example}
\end{align}
Thus, $E_{\text{tot}}^{\text{PoS}}=0.1\,\text{kWh/validator}$, a
97\% reduction compared to PoW\textquoteright s 3.51 kWh/miner, highlighting
PoS\textquoteright s energy efficiency.

\subsection{Proof of Useful Intelligence (PoUI) \label{subsec:PoUI-energy}}

We assume that a PoUI worker operates at high utilization to maximize
computational efficiency. To provide a concrete estimate, we consider
typical values for a GPU server equipped with an NVIDIA A100 GPU,
a common choice for LLM inference. The A100 draws approximately $250\lyxmathsym{\textendash}400\,\text{W}$
depending on its configuration, with system components (e.g., CPU,
memory, cooling) contributing to a total active power consumption,
denoted as $P_{\text{act}}$ of $500\,W$ during full utilization
\cite{nvidia2020a100}. Over one hour, this corresponds to an energy
consumption for useful work of $E_{\text{use}}^{\text{PoUI}}=0.5\,\text{kWh/worker}$.
For security energy, we adopt the staking energy consumption from
Eq. \ref{eq:PoS-energy-example}, where $E_{\text{sec}}^{\text{PoUI}}=0.1\,\text{kWh/validator}$.

When validators also act as workers, they share hardware. The total
energy per node over one hour is modeled as

\begin{equation}
E_{\text{tot}}^{\text{PoUI}}=\kappa_{\text{sec}}E_{\text{sec}}^{\text{PoUI}}+\kappa_{\text{use}}E_{\text{use}}^{\text{PoUI}},
\end{equation}
where $\kappa_{\text{sec}}$ and $\kappa_{\text{use}}$ are weighting
factors representing the proportion of effort allocated to security
(validation) and useful work (task execution), respectively. When
a node focuses solely on validation, $\kappa_{\text{sec}}=1$ and
$\kappa_{\text{use}}=0$, yielding 
\begin{align}
\left|E_{\text{tot}}^{\text{PoUI}}\right|_{\text{val}} & =\left(1\right)\left(0.1\,\text{kWh}\right)+\left(0\right)\left(0.5\,\text{kWh}\right)\nonumber \\
 & =0.1\,\text{kWh}.
\end{align}
When a node performs both task execution and validation concurrently,
$\kappa_{\text{sec}}=1$ and $\kappa_{\text{use}}=1$, and hardware
overlap resulting in:
\begin{align}
\left|E_{\text{tot}}^{\text{PoUI}}\right|_{\text{wrk+val}} & =\left(1\right)\left(0.1\,\text{kWh}\right)+\left(1\right)\left(0.5\,\text{kWh}\right)\nonumber \\
 & =0.6\,\text{kWh}.
\end{align}

\subsection{Comparative Analysis}

\begin{table}
\caption{Energy consumption per node over one hour.\label{tab:energy-consumption}}

\centering{}%
\begin{tabular}{|c|c|c|c|}
\hline 
Mechanism & $E_{\text{sec}}$(kWh/node) & $E_{\text{use}}$(kWh/node) & $E_{\text{tot}}$(kWh/node)\tabularnewline
\hline 
\hline 
PoW & 3.51 & 0 & 3.51\tabularnewline
\hline 
PoS & 0.1 & 0 & 0.1\tabularnewline
\hline 
PoUI & 0.1 & 0.5 & 0.6\tabularnewline
\hline 
\end{tabular}
\end{table}

This section analyzes the energy consumption profiles of PoW, PoS,
and PoUI based on the metrics from Section \ref{subsec:PoW-energy}
- \ref{subsec:PoUI-energy} with a summary provided in Table \ref{tab:energy-consumption}.
\begin{enumerate}
\item Energy Efficiency in Security: PoW consumes the highest security energy
at 3.51 kWh/miner, driven by intensive cryptographic puzzle-solving.
In contrast, PoS and PoUI achieve 0.1 kWh/validator, a 97\% reduction,
due to staking-based mechanisms that eliminate mining.
\item Utility Through Useful Work: PoW and PoS have no useful work energy
($E_{\text{use}}=0$). PoUI allocates 0.5 kWh/worker for AI tasks,
adding value while maintaining efficiency.
\item Total Energy and Trade-Offs: PoW\textquoteright s 3.51 kWh/miner (Table
II) is entirely for security, offering no utility. PoS\textquoteright s
0.1 kWh/validator is the lowest but lacks useful output. PoUI\textquoteright s
0.6 kWh/node, with 0.1 kWh for security and 0.5 kWh for useful work,
achieves an 83\% reduction from PoW while surpassing PoS in utility
(Table \ref{tab:energy-consumption}).
\end{enumerate}

\section{PoUI Market Mechanism \label{sec:PoUI-Market-Mechanism}}

The PoUI consensus mechanism combines useful work with the PoS-like
staking process. In this section, we elaborate the mechanism to regulate
an adequate number of workers in preserving the stability of job market.

\subsection{Dynamic Reward Adjustment}

The PoUI consensus mechanism implements a dynamic reward adjustment
strategy to maintain an optimal balance of workers based on the number
of pending jobs. We assume that job quantities are normalized, such
that one job requires one worker to complete within a single interaction,
denoted as time step $i$. However, jobs of varying complexity, specifically
those requiring multiple interactions, are split into an equivalent
number of normalized jobs to ensure compatibility with this model.

The number of jobs in the market fluctuates over time due to varying
demand. We assume an informed analysis is conducted to derive the
required number of workers in the network, denoted as $\tilde{w}$,
representing the target worker count. The total number of available
workers in the network at time step $i$ is denoted as $w_{i}$, with
the assumption that worker availability is unconstrained. With $\alpha$
denoting the sensitivity of the reward to the disparity between target
and current worker counts, the subsequent equation formalizes a dynamic
adjustment process that modifies the worker reward $r$ to align participation
with job demand, i.e.,

\begin{equation}
r_{i+1}=\begin{cases}
r_{i}\left[1+\alpha\left(\frac{\tilde{w}-w_{i}}{w_{i}}\right)\right] & \text{if }\frac{\left|\tilde{w}-w_{i}\right|}{w_{i}}\geq\Delta\\
r_{i} & \text{otherwise.}
\end{cases}\label{eq:reward}
\end{equation}

Here, the adjustment occurs when the relative difference between the
current and target number of workers exceeds a predefined threshold
$\Delta$, i.e., $\frac{\left|\tilde{w}-w_{i}\right|}{w_{i}}\geq\Delta$.
In such a case, the next reward $r_{i+1}$ increases if the number
of workers is insufficient ($\tilde{w}>w_{i}$), encouraging participation,
or decreases if workers are excessive ($\tilde{w}<w_{i}$), optimizing
resource allocation. If the difference falls below $\Delta$, the
reward remains unchanged, ensuring stability for minor fluctuations.

Simulation is detailed in Subsection \ref{subsec:Simulation-Setup}
to validate this mechanism.

\subsection{Simulation Setup \label{subsec:Simulation-Setup}}

 We assume that the job market requires a target worker count of
$\tilde{w}=250$ to maintain stability. For this simulation, the network
starts with an initial worker count of $w_{0}=100$, and the maximum
number of workers is capped at a sufficiently large value. The dynamic
reward adjustment, formalized in Eq. \ref{eq:reward}, uses parameters
$\alpha=0.2$ (sensitivity to worker disparity) and $\Delta=0.05$
(adjustment threshold). We execute 200 simulation steps to assess
the system\textquoteright s behavior under these conditions.

To model how workers respond to these reward adjustments, we define
their participation dynamics as follow. The number of workers in the
next interaction, denoted as $w_{i+1}$, adapts to reward changes
based on the utility function:

\begin{equation}
w_{i+1}=w_{i}\beta\left(1+\frac{r_{i+1}-r_{i}}{r_{i}}\right)+N\label{eq:workers}
\end{equation}
where $\beta=1$ represents the workers\textquoteright{} sensitivity
to reward variations, and $N=\text{random}\left(-\gamma w_{i+1}^{(\text{c})},\gamma w_{i+1}^{(\text{c})}\right)$
introduces random values as noise, with $\gamma=0.05$ defining the
range of random fluctuations in the worker count. This model assumes
that most workers respond sensitively to reward adjustments, capturing
realistic participation dynamics.

\subsection{Result}

\begin{figure}
\begin{centering}
\includegraphics[width=0.95\columnwidth]{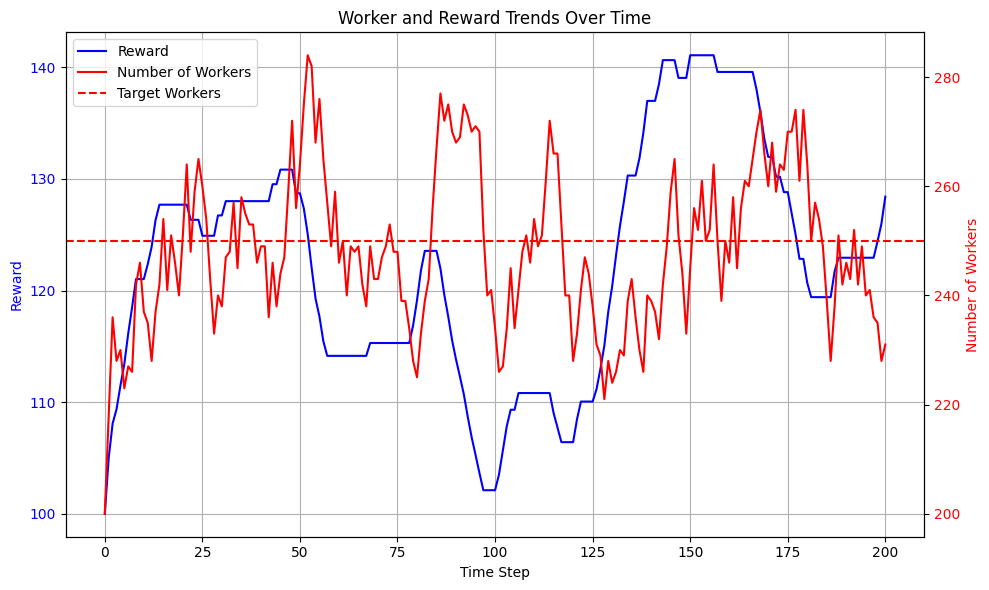}
\par\end{centering}
\caption{Worker and reward trends over 200 time steps in the PoUI simulation,
illustrating the dynamic adjustment of rewards (blue) and the corresponding
number of workers (red) relative to the target worker count of 250
(dashed red line). \label{fig:Worker-and-reward}}

\end{figure}

Fig. \ref{fig:Worker-and-reward} illustrates the dynamic behavior
of the PoUI consensus mechanism over 200 time steps, capturing the
trends in worker rewards (blue line), the number of workers (red line),
and the target worker count (dashed red line at 250 workers). The
simulation evaluates how the dynamic reward adjustment strategy influences
worker participation in response to varying job demands, with parameters
set as $\alpha=0.2$, $\beta=1$, $\gamma=0.05$ and $\Delta=0.05$. 

Initially, the network starts with 100 workers, significantly below
the target of 250. As a result, the reward increases sharply in the
first 25 time steps, peaking at approximately 140, to incentivize
worker participation. This adjustment effectively drives the number
of workers upward, reaching the target around time step 25. However,
the number of workers fluctuates around the target throughout the
simulation, driven by the noise term $N$ in the utility function
\ref{eq:workers}, which models realistic variations in worker participation.

Generally, the reward trend responds to these fluctuations, decreasing
when the worker count exceeds the target (e.g., around time step 50)
to discourage excessive participation, and increasing when the worker
count falls below the target (e.g., around time step 100) to attract
more workers. Despite these oscillations, the system demonstrates
overall stability, as the number of workers generally remains close
to the target of 250, with deviations typically within $\pm20$ workers.
The reward stabilizes around 120 after the initial adjustment phase,
indicating that the dynamic reward mechanism effectively balances
worker participation with job demand.

The result highlights PoUI\textquoteright s ability to adapt to varying
network conditions through its reward adjustment strategy. The mechanism
successfully maintains the worker count near the target, ensuring
efficient job processing while avoiding over- or under-participation.
However, the observed fluctuations suggest that fine-tuning parameters
such as $\alpha$ (sensitivity to worker disparity) could further
reduce variability and enhance system stability in future implementations.

\section{Conclusion \label{sec:Conclusion}}

The PoUI consensus mechanism represents a significant advancement
in blockchain technology by integrating the energy-efficient staking
principles with the productive potential of useful computational work.
Unlike Proof of Work (PoW), which expends substantial energy (e.g.,
3.51 kWh per miner per hour) on security alone, PoUI achieves a dual-purpose
design, securing the network with minimal energy (0.1 kWh per validator)
while dedicating additional resources (0.5 kWh per worker) to run
AI models for valuable tasks. This balance reduces energy waste by
approximately 83\% compared to PoW, while surpassing PoS in utility.
Due to page limitations, several aspects such as the detailed mechanism
of the functional nodes, extended comparisons, broader simulation
scenarios, etc. could not be explored in depth in this paper and will
be revisited in future work.

\bibliographystyle{IEEEtran}
\bibliography{database}

\end{document}